\documentclass{sig-alternate-05-2015}

\usepackage{graphicx}
\usepackage{balance}  
\usepackage{url}
\usepackage{subcaption}
\usepackage{listings}
\usepackage[ruled, linesnumbered]{algorithm2e}
\usepackage{xcolor}
\usepackage{float}
\usepackage{arydshln}

\newtheorem{exmp}{Example}[section]
\newtheorem{mydef}{Definition}
\newtheorem{myproperty}{Property}

\makeatletter
\def\@copyrightspace{\relax}
\makeatother

\begin{document}

\title{SQL Query Completion for Data Exploration}

\numberofauthors{3} 
\author{
\alignauthor
Marie Le Guilly\\
       \affaddr{Univ Lyon, INSA Lyon, LIRIS}\\
       \affaddr{(UMR 5205 CNRS)}\\
       \affaddr{69621 Villeurbanne, France}\\
       \email{marie.le-guilly@insa-lyon.fr}
\alignauthor
Jean-Marc Petit\\
       \affaddr{Univ Lyon, INSA Lyon, LIRIS}\\
       \affaddr{(UMR 5205 CNRS)}\\
       \affaddr{69621 Villeurbanne, France}\\
       \email{jean-marc.petit@insa-lyon.fr}
\alignauthor 
Marian Scuturici\\
       \affaddr{Univ Lyon, INSA Lyon, LIRIS}\\
       \affaddr{(UMR 5205 CNRS)}\\
       \affaddr{69621 Villeurbanne, France}\\
       \email{marian.scuturici@insa-lyon.fr}
}

\date{\today}

\maketitle
\begin{abstract}
Within the big data tsunami,  relational databases and SQL are still there and remain mandatory in most of cases for accessing data.  On the one hand, SQL is easy-to-use by non specialists and allows to identify pertinent initial data at the very beginning of the data exploration process. On the other hand, it is not always so easy to formulate SQL queries: nowadays, it is more and more frequent to have several databases available for one application domain, some of them with hundreds of tables and/or attributes. Identifying the pertinent conditions to select the desired data, or even identifying relevant attributes is far from trivial. 
To make it easier to write SQL queries, we propose the notion of \emph{SQL query completion}:  given a query, it suggests additional conditions to be added to its WHERE clause. This completion is semantic, as it relies on the data from the database, unlike current completion tools that are mostly syntactic.
Since the process can be repeated over and over again -- until the data analyst reaches her data of interest --,  SQL query completion facilitates the exploration of databases. 
SQL query completion has been implemented in a SQL editor on top of a database management system. 
For the evaluation, two questions need to be studied: first, does the completion speed up the writing of SQL queries? Second, is the completion easily adopted by users?
A thorough experiment has been conducted on a group of 70 computer science students divided in two groups (one with the completion and the other one without) to answer those questions. The results are positive and very promising.

\end{abstract}

\section{Introduction}

In the last few years, the database world has not escaped the \textit{Big Data} phenomenon, that contributed to increase tremendously the volume of stored data: it is doubling in size every two years, and should reach 44 zetabyttes ($10^{21}$) by 2020 \cite{rushby2016wiley}. This has thus directly impacted the size of databases, at different levels. Nowadays, it is not unusual to see databases with several hundred of tables, some of them having several hundreds of attributes.
For instance, the database used by the LSST\footnote{\url{http://lsst-web.ncsa.illinois.edu/schema/index.php}} (Large Synoptic Survey Telescope)  contains tables with hundreds of attributes (table \textit{Objects} has 229 attributes for example), and a look at it is convincing to see that writing queries on such schemas are not trivial.
Moreover, names of tables and attributes might not carry meaningful information, rending data accessing particularly difficult. Therefore, it is sometimes hard to find the required information. 

In addition to the increase of data volume and databases schema complexity, more and more people are in contact with databases. As the data is stored in order to be used and to get value out of it, those users are confronted to SQL queries in order to fetch data of interest. Both the data's volume and complexity, combined with the use by non-specialized users, make the query formulation process laborious: as a result, \cite{journals/pvldb/NandiJ11} found that data analysts spend more time on writing a query that the DBMS on processing it.

If more and more people are in contact with data and data\-bases, and are therefore confronted to SQL query formulation, it is also because  data they are trying to access have a \emph{potential value} actionnable in their business. Analysts explore data in order to understand it, and to gain additional knowledge through this process. Data exploration tools are therefore crucial as they help users to go to interesting regions of their dataspace, and to identify relevant information or patterns in the data. 

In databases, the first step of an exploration consists or is equivalent to the formulation of a SQL query, as it is a way to reach a specific set of data. Moreover, the conditions specified in such a query are already a way to characterize the data. This process is somehow related to web search engines that are used daily to reach specific websites, documents, etc, through keywords to describe what to look for. In this context, the task is easier than with databases from a user point of view: the language to express the query is not constrained, the syntax can be a bit loose \ldots and more importantly, some help is provided to them: once the beginning of the search is typed in, the search engines suggests completions to the beginning of an input. These completions guide the search, by refining the idea, or by  indicating keywords a user had not thought about. Such completions are based on the data indexed by the search engine, and also on user's history (previous researches, similar queries by other users...). 

\textbf{Problem statement}
To make it easier to write SQL queries, we propose the notion of \emph{SQL query completion}:  
\begin{quote}
\emph{Given an initial SQL query $Q$ against a database $d$, suggest additional conditions to be added to the WHERE clause of $Q$ such that the new query identifies a meaningful subset of the data identified by $Q$}. 
\end{quote}

This semantic SQL query completion allows the exploration of databases, since the process can be iteratively repeated over and over again until the data analyst reaches her data of interest.
To the best of our knowledge, such a form of SQL query completion has not been studied yet. In databases, some help is provided for query formulation, but it is mostly syntactic:  SQL editors of leading database management systems propose auto-completion of keywords from the SQL Language, as well as dedicated SQL editors like \emph{SQL Complete}\footnote{\url{https://www.devart.com/dbforge/sql/sqlcomplete/}} or \emph{SQL Prompt}\footnote{\url{https://www.red-gate.com/products/sql-development/sql-prompt/}}. 
Basically, they fetch information for the database's schema to provide auto completion of tables and attributes names, which are a very simple form of semantic completion. However, this completion is only based on the schema, and surprisingly, does not look at all at the data contained in the database. Therefore the help offered by those completion systems is limited, as it does not contain any additional intelligence regarding the content of the database on which the query is evaluated. They only accelerate the process by suggesting information that an analyst would have to search in the schema or in a SQL handbook otherwise, and spare the user some manual writing.

There are also some works that have been done using user's history to suggests new queries. Such a solution is exposed in \cite{conf/cidr/CetintemelCDDDKPZ13}, to suggest queries based on previous queries asked by similar users. However, this is limited to presenting an already existing query. 

\begin{table}
\begin{center}
\centering
\begin{tabular}{ccccc}
\hline
EmpNo & LastName & Gender & Salary & Commission \\
\hline
e10 & SPEN & F & 41160 & 1300 \\
e20 & THOMP & M & 41250 & 7400 \\
e30 & KWAN & F & 39850 & 5200 \\
e40 & SMITH & F & 40525 & 1400 \\
e50 & GEYER & M & 40175 & 1100 \\
e60 & STERN & M & 39560 & 6200 \\
e70 & PULASKI & F & 40120 & 800 \\
e80 & FREY & M & 40625 & 6600\\
e90 & HENDER & F & 39450 & 6700 \\
e100 & SPEN & M & 41560 & 900 \\
\hline
\end{tabular}
\caption{Running example: Employees dataset}
\label{tab:example}
\end{center}
\end{table}

\begin{exmp}
\label{exmp:completions}

Let us assume that Alice, a data analyst, has access to the database of a company, containing data about employees, presented in table \ref{tab:example}. She is
asked to find information that can be valuable for the company in terms of discriminatory behaviour for female employees. After looking at some data samples, her first idea is to look for a correlation between the gender and salary of employees.
Without any other intuition, she can start with the following simple query $Q$:

{\normalfont
\begin{lstlisting}[mathescape=true]
Select Gender, Salary
From Employees
\end{lstlisting}
}

Although useful, it does not reveal anything to Alice.

With the semantic completion proposed in this paper, Alice would have the choice to get help -- without any new interaction with the system -- from the three following completions of her initial query:

{\normalfont
\begin{lstlisting}[mathescape=true]
Select Gender, Salary
From Employees
Where commission $\geq$ 6200
\end{lstlisting}
\begin{center}
Completion 1 (returns 4 tuples)
\end{center}

\begin{center}
\begin{lstlisting}[mathescape=true]
Select Gender, Salary
From Employees
Where commission < 6200
  and sex = 'F'
\end{lstlisting}
Completion 2 (returns 4 tuples)
\end{center}

\begin{center}
\begin{lstlisting}[mathescape=true]
Select Gender, Salary
From Employees
Where commission < 6200
  and sex $\neq$ 'F'
\end{lstlisting}
Completion 3 (returns 2 tuples) \\
\end{center}
}
\end{exmp}

These completions are relevant for Alice: the first one selects employees with a high commission. The two others select employees with a lower commission, among which they discriminate between male and female employees. And it seems that they are more women with low commission than men, which could indicate a discrimination in term of commission based on the gender of employees.

Therefore, those completions are a way to highlight an information and patterns which could be pertinent for Alice, but that would have been hard, or even impossible to find by just looking at the data.

Moreover, the \texttt{commission} attribute appears in completions, even though it was not considered pertinent enough by Alice in her initial query. It helps her to see that the discrimination might not be done on the salary as expected, but on the commission of female employees.

The above example is also useful to explain what a semantic query completion could be, and how it could be useful. Indeed, the following observations can be made:

\begin{itemize}
\item $Q$ is contained in each of its completions, but they have additional conditions in their WHERE clause;
\item They all lead to different result sets, exploring a different subspace of the initial query's result set. There is no new tuple that did not appear in the evaluation of $Q$, the completion is here a way to narrow down the initial result set;
\item Each completion captures a different pattern in SQL, potentially meaningful;
\item The completion is  a way to draw attention on new attributes that might have been unnoticed. 
\end{itemize}

This example gives the intuition of the type of SQL query completion proposed in this paper. \\

\textbf{Paper contribution}
To the best of our knowledge, we study for the first time the problem of SQL query completion based on the semantic of the query to be completed. We have made the following contributions:
\begin{itemize}
\item A definition of a class of SQL query completion ;
\item Given a database, a query and the number of desired completions, an algorithm to compute such completions that does not require any boring user input.
The idea is to identify groups of similar tuples in the answer set of an initial query, and based on those groups, to build decision clauses to discriminate between them. These decision clauses are then injected into the initial query in order to build new queries, that we call completions;
\item An implementation of this algorithm in a SQL editor prototype;
\item An experiment on a sample group of 70 computer science students, measuring the usefulness and the acceptability of such completions.
\end{itemize}

\textbf{Paper organization}
Section \ref{sec:prelim}  introduces the preliminaries and the definitions of SQL query completion. Section \ref{sec:algo} exposes a solution to compute such completions. Then section \ref{sec:expe}  presents an implementation, and experimentations that were conducted to evaluate the completion in SQL. Finally section \ref{sec:state} summarises  the related work, before concluding in section \ref{sec:ccl}.

\section{SQL queries and their completions}\label{sec:prelim}

\subsection{Preliminaries}
Let us start by introducing the basic notations to be used throughout this paper. We assume the reader is familiar with databases notations (see \cite{Levene:1999:GTR:553537} for details). 
Let $\mathcal{D}$ be a set of constant and $\mathcal{U}$ a set of attributes. We consider a database $d = \{r_1,r_2,..., r_n\}$ over a database schema $R = \{R_1, R_2, ... R_n\}$, where $r_i$ is a relation over a relation schema $R_i$ and $R_i \subseteq  \mathcal{U}$, $i \in 1..n$. 
We consider the SQL and the relational algebra query languages without any restriction. We will switch between both languages when clear from context. A query $Q$ is defined on a database schema $R$ and $ans(Q,d)$ is the result of the evaluation of $Q$ against $d$. In the sequel, to define the completion of any query $Q$, we will use two operators: $\pi_X$ the projection defined as usual with $X \subseteq \mathcal{U}$, and $\sigma_F$ the selection, where $F$ is a conjunction of atomic formulas of the form $A \theta B$ or $A \theta v$, with $A,B \in \mathcal{U}$, $v \in \mathcal{D}$ and $\theta$ a binary operator in operation in the set $\{<,>,\leq,\geq,=,\neq\}$

\subsection{Query completion}
The simple question we have to answer is:
\textit{How to define a completion of a query $Q$ ?}

Different kinds of completions can be imagined, and many different ways to compute them. However, as well as for search engine, a query's completion should help to narrow down the results. Therefore, the intuition is that
if $Q_{cmp}$ is a completion of a query $Q$ then $ans(Q_{cmp}, d) \subseteq ans(Q, d)$.
However, many queries could comply with this property
, some that might not be  considered as \textit{completions} of $Q$.

\begin{exmp}
\label{exmp:pas_comp}
Let $Q_1$ and $Q_2$ be the two queries over database presented in table \ref{tab:example}.
\end{exmp}

\begin{minipage}{.4\linewidth}
\begin{lstlisting}[mathescape=true]
$Q_1$:
Select EmpNo
From Employees
Where sex = 'F'
\end{lstlisting}
\end{minipage}
\begin{minipage}{.6\linewidth}
\begin{lstlisting}[mathescape=true]
$Q_2$:
Select EmpNo
From Employees
Where Commission = 1300
\end{lstlisting}
\end{minipage}

\textit{The result of $Q_2$ is contained in the result set of $Q_1$ but $Q_2$ is not a completion of $Q_1$
}

To define query completion,  additional conditions should be given at the syntactic level. This requires to formalize this relationship between a query and its completions.

\begin{mydef}
\label{def:completion}
The completion $Q_{cmp}$ of $Q$ is defined by:
$$ Q_{cmp} = \sigma_{c_1 \wedge ...  \wedge c_n}(Q) $$
where $c_i$ is an atomic formula, for every $i \in 1..n$

\end{mydef}

Specified like this, the number of queries that could be considered as completions is still infinite and it seems difficult, if not impossible, to give a meaningful definition of a single completion.

For this reason, we do not consider a single completion of a given query, but a set of $k$ completions such that global properties can be defined, especially:
\begin{itemize}
\item The union of completions set of size $k$ is equal to the initial data set represented by the initial query ; and
\item Each completion is as much as possible dissimilar to each other.
\end{itemize}

We come up with the following definition.

\begin{mydef}
\label{def:completionset}
A \emph{k-set completion} of $Q$, denoted by $C_Q$, is defined as:  $C_q= \{ Q_1,Q_2,..., Q_k \} $ such that: 
\begin{itemize}
\item $Q_i$ is a completion of $Q$, for all $i \in 1..k$

\item $ans(Q_i,d) \cap ans(Q_j,d) = \emptyset$, , for all $i,j \in 1..k, i\not= j$
\item $\bigcup\limits_{i=1}^{k} ans(Q_i,d) = ans(Q,d)$
\end{itemize}
\end{mydef}

Clearly, a $k$-set completion forms a partition of $ans(Q,d)$. Those restrictions provide a nice setting to see every completion as a good candidate to start a data exploration process.

\section{Computation of k-set completion using machine learning}
\label{sec:algo}

We argue that computing k-set completions is a new problem, even if many subproblems have been studied for years in different communities, for example clustering in machine learning, \emph{query reverse engineering} in database \cite{Tran:2014:QRE:2673202.2673266}  or \emph{redesciption mining} in data mining \cite{parida2005redescription}. 
In the sequel, we explain the solution based on a two-steps process.

\subsection{Division of tuples via clustering}
Taking into account definition \ref{def:completionset} and the need for non similar answer sets for each completion, it is obvious that the division of the tuples in $ans(Q,d)$ should not be random. The $k$ sets obtained through this division have to be pairwise disjunct, and cover $ans(Q,d)$ entirely. Moreover, they should be as different as possible. Clustering algorithms (see \cite{Han:2005:DMC:1076797} for an overview) offer a nice setting to provide effective approximate solutions.

In this paper, we focus ourselves on the k-means algorithm \cite{Lloyd:2006:LSQ:2263356.2269955} since the parameter $k$ is part of the input. This algorithm requires to be able to compute a distance between a given pair of tuples, which is feasible but sometimes tricky. More elaborated techniques is left for future work.

After the clustering, each tuple can be assigned to a single cluster, providing an unique opportunity to simplify drastically the role of the user in the boring task of tuple labelling. However, this label is only here for background completion computation, and is therefore not shown to the user.

Technically, a new attribute, called \texttt{cluster}, is added to the schema of the query to keep track of the cluster corresponding to each tuple.

\begin{exmp}
Table \ref{tab:cluster} presents tuples from table \ref{tab:example}, with the additional attribute \textit{cluster} that is the cluster tuples have been assigned to. 

\begin{table}
\begin{center}
\centering
\begin{tabular}{ccccc:c}
\hline
EmpNo & LastName & Sex & Salary & Commission & \textit{Cluster}\\
\hline
e10 & SPEN & F & 41160 & 1300 & \textit{2} \\
e20 & THOMP & M & 41250 & 7400 & \textit{1}\\
e30 & KWAN & F & 39850 & 5200 & \textit{2} \\
e40 & SMITH & F & 40525 & 1400 & \textit{2} \\
e50 & GEYER & M & 40175 & 1100 & \textit{3} \\
e60 & STERN & M & 39560 & 6200 & \textit{1} \\
e70 & PULASKI & F & 40120 & 800 & \textit{2} \\
e80 & FREY & M & 40625 & 6600 & \textit{1}\\
e90 & HENDER & F & 39450 & 6700 & \textit{1} \\
e100 & SPEN & M & 41560 & 900 & \textit{3} \\
\hline
\end{tabular}
\caption{Employees dataset labelled by clustering}
\label{tab:cluster}
\end{center}
\end{table}
\end{exmp}

\subsection{Construction of completions with a binary decision tree}
The clustering addresses the first step of the solution. The second step is now to find at least one query to describe each of the clusters, if possible.  It should be noted that in our approach, the clustering is done to offer an automatic labelling of tuples, based on a guided division of data, to avoid a manual labelling, and to reveal underlying groups of data. The decision tree will therefore try to discriminate between clusters, and even though the use of a clustering algorithm should facilitate the finding of meaningful discriminating clauses, the purpose is not do describe perfectly each cluster. We rather propose to use the clusters as "clever" labels.

Using decision trees to generate SQL queries is a technique that has already been exploited \cite{cumin2017data}. To be able to reach the objectives defined for our SQL completions, we follow the same path but with \emph{binary decision trees} (BDT) \cite{breiman1984classification}, which is a tree splitting at each node on exactly two opposite conditions. In the sequel, we introduce an approach based on 1) the computation of a constrained BDT with $k$ leaves from the data partition obtained before, and 2) the transformation of the  BDT into SQL statements.

\subsubsection{Obtaining a constrained BDT with $k$ leaves from a given data partition}

Clearly, we do not need to determine a full BDT since we just have to output $k$ leaves from it, each leaf giving rise to one SQL completion. We assume our data allows the construction of a tree with at least $k$ leaves. Constrained generation of BDT given a specific number of leaves has been studied in \cite{wu2016decision}. In our case, we just need to explore levelwise the search space (breadth-first search) and stop as soon as the number of leaves exceeds $k$.

Indeed, the depth of the BDT is bounded by $\lceil \log_2(k) \rceil$ and $k -1$. 
Both bounds are atteignable: the first one with a \emph{full binary tree} and the second one with a \emph{right deep tree}.

This optimisation turns out to be very efficient in practice.

\begin{figure}
\centering
\includegraphics[width=\linewidth]{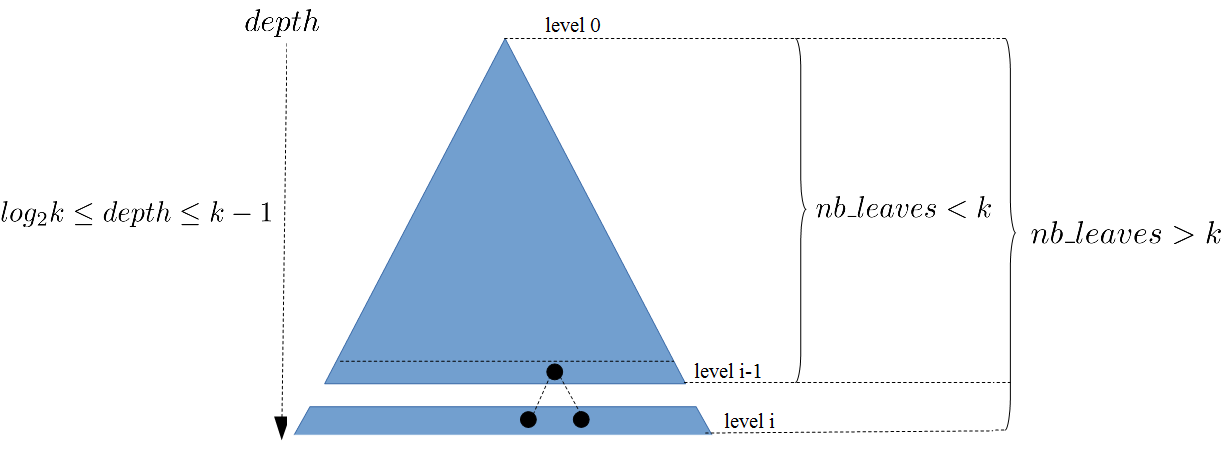}
\caption{Construction of a binary decision tree given a fixed number of leaves}
\label{fig:leaves}
\end{figure}

To reach exactly the $k$ leaves constraint, let us consider the figure \ref{fig:leaves}: assume the number of leaves int the BDT is less than $k$ a level $i-1$, and greater or equal to $k$ at level $k$. If the number of leaves at level $i$ is equal to $k$, then stop. Otherwise, while the number of leaves remains greater than $k$, replace two leaves at level $i$ from the same parent by turning this parent into a leaf at level $i-1$ (using a majority vote to assign a class to this new leaf).

\begin{exmp}
From the clustering in table \ref{tab:cluster}, the binary decision tree of figure \ref{fig:tree} can be obtained. In this running example, the decision tree leaves matches exactly with the clusters. 
\begin{figure}
\centering
\includegraphics[width=.9\linewidth]{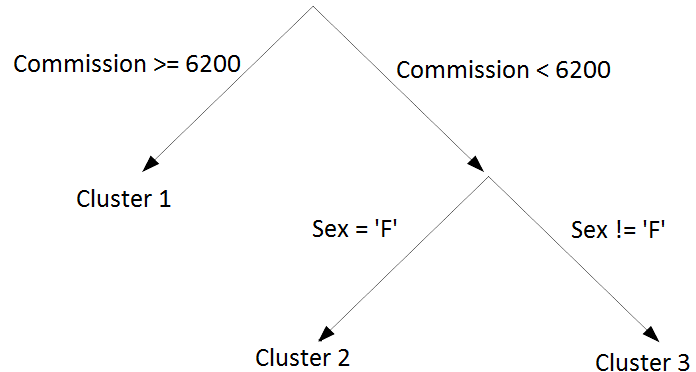}
\caption{Binary decision tree from Table 1}
\label{fig:tree}
\end{figure}
\end{exmp}

The previous example points out that the clustering and the binary decision tree may coincide. However, this is not true in general since some tuples may fall into the wrong cluster or some clusters could be lost by the binary decision tree. 
In a data exploration process, we argue this is not a real issue: the new queries proposed by our techniques are a first step to help the analyst to write her query.

\subsubsection{Obtaining SQL statements from a BDT with $k$ leaves}

Once the binary tree has been constructed, each leaf can be reached through a unique decision path. The decision path from the root of the tree to a specific node can be written as the conjunction of each decision encountered along the road. This is why it is so convenient to go from a decision tree to a SQL query. After exploring all the path in the tree, each conjunction can be directly injected in the \textit{where} clause of an SQL query, and therefore give a new completion.

\begin{exmp}
From the decision tree on figure \ref{fig:tree}, every completion from example \ref{exmp:completions} can be obtained easily. 
\end{exmp}

\subsection{Algorithm proposal}
In order to combine the two steps described previously, and to specify how the completion set is to be computed, the algorithm \ref{algo} is proposed hereafter. 

\begin{algorithm}
    \SetKwInOut{Input}{Input}
    \SetKwInOut{Output}{Output}

    \underline{procedure Completion} $(Q,d,k)$\;
    \Input{A query $Q$ over $R$, \\ $d$ a database over $R$, \\ $k$ the number of completions} 
    \Output{$S_c$ a set of k completions of $Q$}
    	
	\If{$Q = \pi_X(Q')$ 	 \tcp*{remove the projection} } 
		{Q = Q'} 
	
	wd = $ans(Q,d)$ \tcp*{wd: working data}

    lwd $=kmeans($wd, k$)$ \tcp*{lwd: labelled wd} 
    
    treeWithKLeaves $= LevelwiseBDT($lwd, k$)$\\
    
	conjunctions $= getRules($treeWithKLeaves$)$\\
	
	$S_c$ $= \{\}$
	
	\ForEach{c $in$ conjunctions}{%
      $S_c = S_c \cup \sigma_c(Q)$  
            
    }    
    
    \KwRet{$S_c$}\;
    
    \caption{Query completion procedure}
    \label{algo}
\end{algorithm}

Lines 2-5, we identify first whether the outermost operator is a projection or not. In such a case, we remove it to 
augment the chances of clustering on pertinent attributes. Indeed, data analysts do not necessarily select the attributes that are the most discriminant or the most interesting to learn on. Therefore, when evaluating the initial query, the returned attributes might not be the most useful to compute completions.

This first step consists in opening the query to as many attributes as possible within the tables concerned by the query. We have proposed a simple but effective solution, but there is still room for improvement, more attributes could be added or some attributes could be removed, especially those (string or categorical) with a very large number of different values.

Line 6, we compute the result set of the initial query. Clearly, if the size of the result is expected to be large, some restriction could be added to save time and ressources, for instance by limiting the output to the first 10 000 or 100 000 tuples or by using sampling techniques. \\
Line 7, the clustering transforms the dataset $wd$ into a labelled dataset $lwd$ in which each tuple is labelled with the cluster it was assigned to. \\

Line 8, a constrained BDT with $k$ leaves is computed from the labelled dataset such that its number of leaves is equal to the $k$ value.
In other words, the BDT is built such that its depth is as small as possible while the number of leaves has to be equal to $k$, as explained in the previous section.

Lines 9-12 this procedure produces exactly $k$ completions: for each leaf of the tree, the conjunction of clauses leading to it from the root is computed. 
This conjunction is then added to the selection clause of $Q$, creating a new completion. 

The conditions from definition \ref{def:completionset} are satisfied by the proposed algorithm as stated in the following property.
 
\begin{myproperty}
Let $d$ a database over $R$, $Q$ a query over $R$, and $k$ an integer.
$completion(Q,d,k)$ is a k-set completion of $Q$, i.e for $\{Q_1, Q_2,... Q_k\}$ in $completion(Q,d,k)$, and for all $i,j \in 1..k$, $i\not= j$:
\center
\begin{equation}
Q_i \mbox{ is a completion of } Q 
\end{equation}
\begin{equation}
ans(Q_i,d) \cap ans(Q_j,d) = \emptyset
\end{equation}
\begin{equation}
\bigcup\limits_{i=1}^{k} ans(Q_i,d) = ans(Q,d)
\end{equation}
\end{myproperty}

\begin{proof}
(1) By construction, and with respect to definition \ref{def:completion}, $Q_i$ is a completion of $Q$

(2) At each node of the binary decision tree, there is one split leading to the creation of two child nodes. This split is done on one attribute $A$, for one threshold value $t$ if $A$ is numeric or a value $v$ otherwise. The first child node takes all tuples in the dataset for which $A \leq t$ (respectively $A = v$), the second child node takes the rest, i.e tuples for which $A > t$ (respectively $A\not= v$)\footnote{In case of null values on $A$,  one of the two conditions of a split should integrate a test of the form $A$ \texttt{IS NULL},  not to miss any tuple.}. As a consequence, any tuple in the dataset reaches one and only one node at each split, and therefore one and only one leaf of the tree. As each completion corresponds to one leaf, they all contain different tuples from $ans(Q,d)$. 

(3) By property of completion we have $ans(Q_i, d) \subseteq ans(Q, d)$, so we obtain $\bigcup\limits_{i=1}^{n} ans(Q_i,d) \subseteq ans(Q,d)$. Moreover, as a node splits on opposite conditions, any tuple from $ans(Q,d$) satisfies one and only one condition at each split. Therefore any tuple $t$ from $ans(Q,d)$ necessarily ends up in the result set of a completion, so there exists $i \in 1..n$ such that $t \in ans(Q_i,d)$ and therefore $\bigcup\limits_{i=1}^{n} ans(Q_i,d) \supseteq ans(Q,d)$.
\end{proof}

Clearly, we make use of machine learning algorithms that are currently well-known when solving various classification or prediction tasks. 
Usually those algorithms are fine-tuned to be perfectly fitted to the task at hand, requiring time and background knowledge from a data analyst \cite{Han:2005:DMC:1076797}. 
In our context, this is not possible, but
optimal performance is not crucial. 

The better the algorithm, the better the completions, but even with the double approximation from the clustering and the decision tree, the information from the completion set remains definitely useful as shown in the experiments.

\section{Implementation and Experimentations}
\label{sec:expe}

\subsection{Algorithm implementation}
Algorithm \ref{algo} was implemented using Python 3 and using \textit{SQLite}\footnote{\url{https://www.sqlite.org/}} as DBMS. For sake of simplification, the implementation is for now limited to numerical attributes. The \textit{kmeans} algorithm is taken from the \textit{scikit-learn} \cite{scikit-learn} library. The binary decision tree classifier is also from this library, based on the CART algorithm \cite{cart84}, but adapted to comply with the constraint of the number of leaves.

\subsection{SQL Editor Prototype}
\begin{figure*}
\centering
\includegraphics[width=\textwidth]{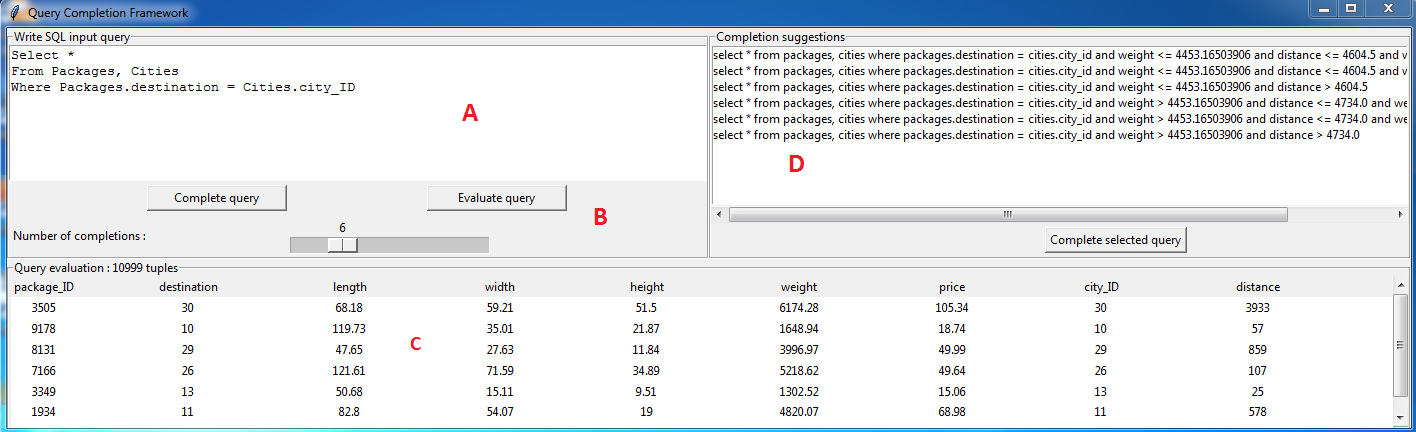}
\caption{SQL query completion prototype}
\label{fig:ihm}
\end{figure*}

We also implemented a SQL editor Prototype, with a basic user interface, that in addition to basic SQL functionalities, offered the possibility of query completion.
Also developed in Python, this interface consists in four distinct zones (see figure \ref{fig:ihm}):

\begin{itemize}
\item A text field to write the  query to be completed (A)
\item A parameter field to specify the maximum number of completions to return (B)
\item A zone to display the answer set of a query (C)
\item A zone where the completion set of the input query is presented once computed (D)
\end{itemize}

\subsection{Experimentations}
The objectives of experimentation was to prove the utility of completion as presented in this paper. It was decided to explore two different categories of measures :

\begin{itemize}
\item How does the completion improves the query writing process: in terms of writing time, is it faster to write a query using completion ?
\item How well can the completion be accepted by users ? Is it easy to use ? How well do users adapt to it ?
\end{itemize}

\subsubsection{Organization}

A group of 70 computer science student (last year bachelor students and master students) was gathered. They all had at least basic knowledge in SQL and data management. They all agreed to participate in a one hour experimentation that was meant to test the completion tool. They were initially only told that they would have to adress several SQL-related challenges.

Prior to the experiment, participants were randomly divided into two groups. The division was however balanced in terms of number of students from each level (bachelor, first year and second year master students). The experiment required to evaluate SQL queries on a database. For this, the first group (referred to as group CMP from now on) had access to the completion tool, while the other (referred to as group NoCMP from now on) had a tool that was designed to be similar to the one of group CMP, but without completion. This disposition was chosen to be able to compare the results of the two groups, i.e to see the difference between groups with and without completion, while working under similar conditions (softwares with similar functionalities in terms of classic querying tools).

Each group was asked the same ten questions on a database\footnote{\emph{Note to the reviewers}: the DB schema and corresponding questions are available in appendix at your discretion. All experimentation material is available online at \url{https://marielgy.github.io/sql_experimentation/}}. 

\subsubsection{Design of the test}
When conceiving the questions, our purpose was to propose a fair situation for groups CMP and NoCMP. For this reason, we eliminated several types of questions :

\begin{itemize}
\item Questions that were trivial with completion, but impossible to do without completion
\item Questions for which completion has no interest: queries with empty result sets, dates comparison, specific operators from DBMS...
\end{itemize}

All questions exposed a scenario, and then asked to find out the SQL query to solve the scenario. The questions were ordered from the easiest to the most difficult, and separated into two categories :

\begin{itemize}
\item The first three queries were classic SQL queries, that are typical of SQL lessons for beginners: they were questions directly and easily transformable into SQL queries, on which the completion tool was not useful. Those questions allowed to verify that each participant really had basic SQL skills. It was also a way to verify that group CMP and NoComp had similar results, and were therefore well balanced.
\item The other questions (number 4 to 10) were designed to be more open-ended, in the sense that the conversion of the question into an SQL query was not straightforward. 
The purpose was to mimic the kind of loosely-specified questions a data analyst is confronted to when exploring unknown data, or answering questions from non-SQL users.
The specification of questions was less strict, as selection conditions were not specified in terms of numbers, but using adjectives such as \textit{higher, bigger, lower, above average, low, etc}. However, the final SQL query required to identify numerical conditions to discriminate between the tuples, and to translate the description of data given in the question.
But even with those specifications, the level of difficulty for the second group of questions was not easy to settle: formulating  those \textit{vague} questions requires to choose carefully the vocabulary used in the question. 
\end{itemize}

To help the NoCMP group, and make sure they would still be able to complete the test, we added to types of information. First, we provided the number of tuples each query was supposed to return, to provide an indication.
Second, for some questions, we proposed data visualizations and asked participants to formulate queries that would return specific datapoints of these visualizations. This was a way to ask participant to transform a visual pattern into a query, so they had to identify the pertinent conditions to characterize the given pattern.

Participants had one hour to answer the 10 questions. They had to read the question, use the tool to write queries and evaluate them on the database, and once they thought they had the right query, they had to submit it online. They were not told whether their answer was right or not: in a real life scenario,  analysts have to know themselves if they reach their desired data or not. During the hour of experimentation, we were able to monitor the time each participant spent on each questions. After the experiment, we also checked whether the answers they submitted were correct or not. Moreover, we were able to say, for each question, if participants from group CMP had used completion or not to generate their submitted query.

At the same time, group CMP had to deal with an additional difficulty, as they had to handle the completion tool. Indeed, they had never used it before, and they did not receive any specific formation on how to use the tool before the test. They were only given a one-page instruction sheet on how completion worked (see website \footnote{\url{https://marielgy.github.io/sql_experimentation/}}). But they did not get any additional time, and had to use the hour to both answer the questions and master the completion (even though they were not forced to use it). This was done to avoid influencing them on their use of the completion, and to see how they would adapt to this new functionality.

Finally, after completing the 10 questions, participants were asked to answer a quick survey to collect their opinion and feelings on the experiment.

\subsubsection{Results}

\begin{figure}
\centering
\includegraphics[width=.9\linewidth]{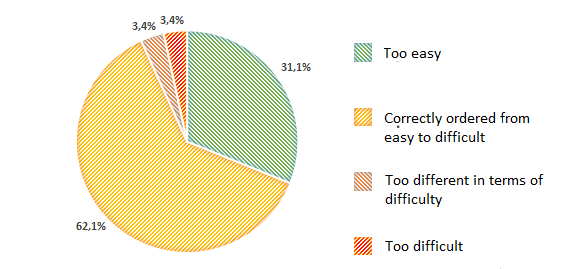}
\caption{Difficulty of questions perceived by group CMP participants}
\label{fig:adifficult}
\end{figure}

\begin{figure}
\centering
\includegraphics[width=.9\linewidth]{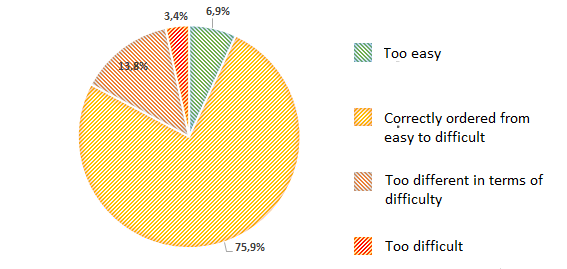}
\caption{Difficulty of questions perceived by group NoCMP participants}
\label{fig:bdifficult}
\end{figure}

\paragraph{Validation of experimental setup}
First, we validated that the constraint specified for question formulation.
The objectives were fulfilled regarding the difficulty of questions as well as equity between groups. On figure \ref{fig:adifficult} and \ref{fig:bdifficult}, we can see how participants felt regarding the difficulty of questions. In both groups, only 3.4\% of participants felt like questions were too difficult. In both case, the majority of participants felt like questions were correctly ordered from easiest to more difficult. The only difference is that more people in group CMP felt like questions were too easy (31.1\% against 6.9\%), as the completion tool helped them in answering the questions that were supposed to be really difficult.

On this first evaluation, we met our objectives with respect to our questions difficulty. This is also an indication that query completion can make answering SQL questions easier for users.


\begin{figure}
\centering
\includegraphics[width=\linewidth]{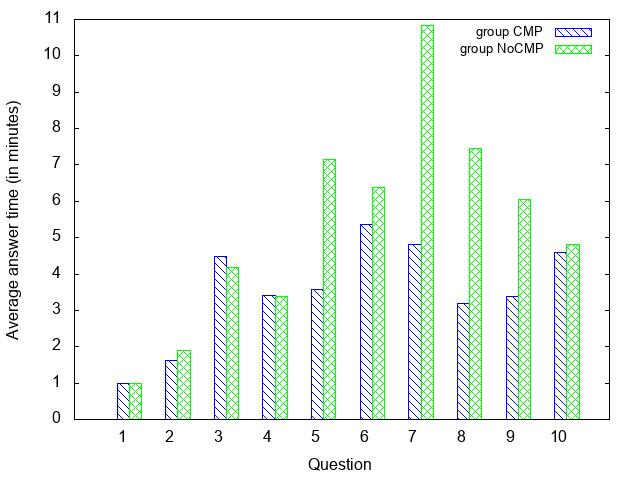}
\caption{Histogram of average answering time per question, for group CMP and group NoCMP}
\label{fig:g1}
\end{figure}

\paragraph{Completion and query writing time}
To analyze the impact of completion regarding query writing time, the first result that is interesting to look at is how much time did each group spend on average answering each question: those results are presented on figure \ref{fig:g1}. There are several interesting points to notice on this figure :
\begin{itemize}
\item For questions 1 to 3, the results of the two groups are similar, which was actually the initial objective. When completion was not necessary, the performance of both groups were equivalent.
\item Question 4 was easy for both group, as could be expected as the visualization was here to help .
Even though completion could have helped on this question, it does not seem to have made a difference, as the average answering time is very similar for both groups. This means that completion is not necessary if a good visualisation is available to the user.
\item For questions 5 to 10, the difference between the two groups is much more important and it seems clear that group CMP performed considerably faster than group NoCMP. This is a strong argument to support the fact that SQL query completion can indeed make the SQL query writing faster. The difference is stronger for questions 7 and 8, which  seem to have been the most difficult questions for participants.
\end{itemize}

\begin{figure}
\centering
\includegraphics[width=\linewidth]{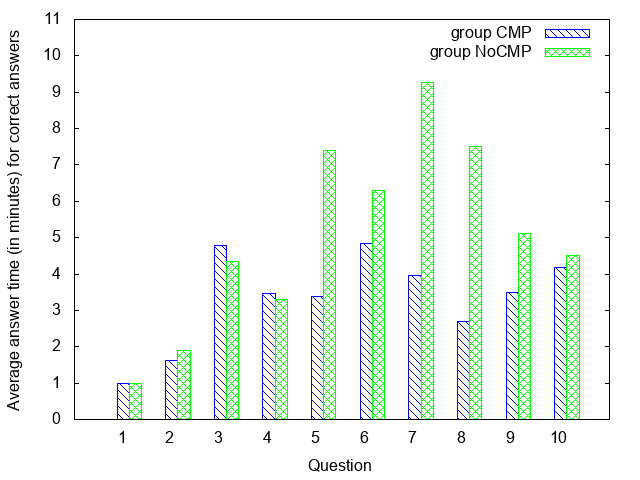}
\caption{Histogram of average answering time per group, for each question, only for correct answers}
\label{fig:g3}
\end{figure}

However, figure \ref{fig:g1} takes into account all answers from participants, which means that some of those answers might be wrong. And a participant who did not gave a good answer might have spent a lot of time on a question looking for the answer without finding it, or on the contrary given up quickly as he did not know how to fnd the answer. For this reason, figure \ref{fig:g1} was recomputed, taking into account only the answering time from participants who had given the correct answer for the considered question: such results are presented on figure \ref{fig:g3}. The tendency is similar, and group CMP still performs considerably faster than group NoCMP. Actually, results from group CMP are even slightly better, especially for more complex queries like for question 7 and 8.

\begin{figure}
\centering
\includegraphics[width=\linewidth]{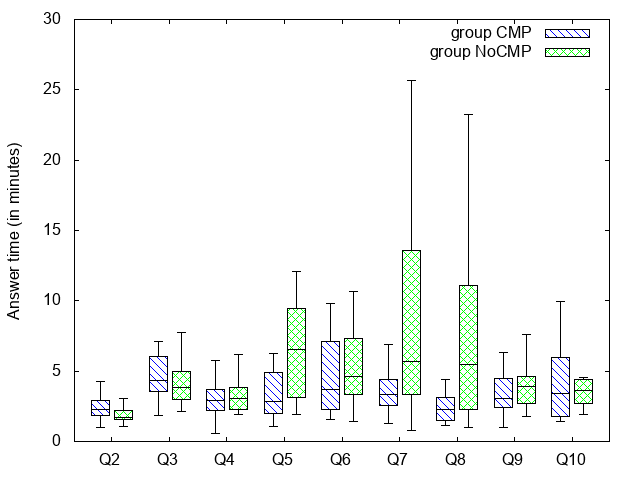}
\caption{Boxplot of answering time per question, for groups CMP and NoCMP, only for correct answers}
\label{fig:boxplot}
\end{figure}

To understand the behaviour of participants, it is possible to look at the boxplot of answering time per question for each group, on figure \ref{fig:boxplot}, which also take into account only correct answers. The main observation is that results of group CMP are much more packed than for group NoCMP: participants who had access to completion had a way to help them if they were stuck on a question, contrary to group NoCMP participant who had to search by themselves until they identified the answer. This is flagrant once again for question 7, where someone spent more than 25 minutes looking for the answer. 

On this second evaluation, our initial objectives are completed: when evaluated in similar conditions, the group with access to completion performed faster than the group with only classic SQL tools.

\begin{figure}
\centering
\includegraphics[width=\linewidth]{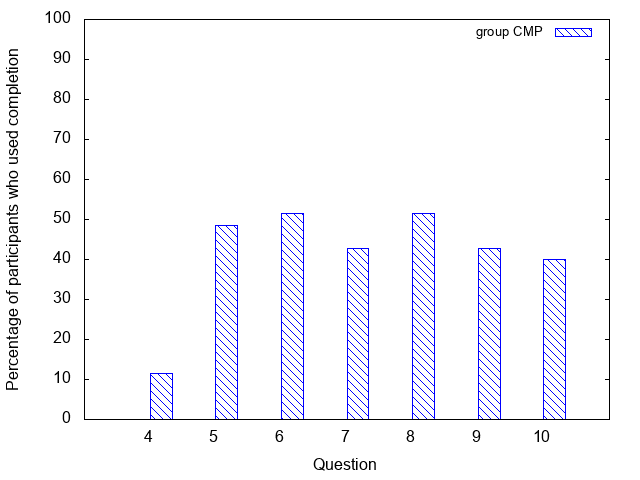}
\caption{Percentage of participants from group CMP who used completion to answer the question, for questions 4 to 10}
\label{fig:g4}
\end{figure}

\paragraph{Completion acceptance}
As mentioned previously, it was also possible to say whether a participant had used completion for a given question or not. The proportion of completion use per question is presented on figure \ref{fig:g4}. We only presented question 4 to 10 on which completion was possible.
It can be seen on this figure that participants did not always use the completion tool. In total, 70\% of participants used completion at least once, while the others completed the test without using it.

\begin{figure}
\centering
\includegraphics[width=\linewidth]{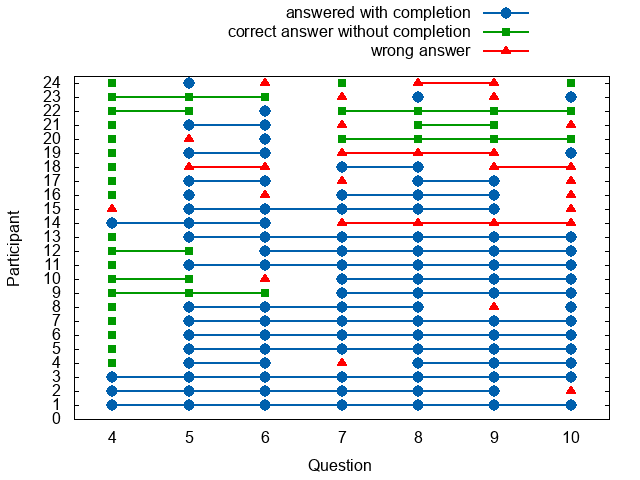}
\caption{Type of answer per question, for participants who used completion during the test}
\label{fig:g8}
\end{figure}

Additional results were analysed to understand those observations. We first analysed, the way participants had used completion: on figure \ref{fig:g8}, interesting patterns can be observed. The main observation to do is that once participants have used completion for a question, they are very likely to use again in the next question. This is indicated by the continuous blue lines on this figure. This is a really important result, as it showed that once a user has understood the utility of completion, he will use it again. This observation is particularly true for participants number 1 to 13, which in addition did not make many mistakes. Participants 14 to 19 also used completion a lot after their first use, but made more mistakes: when looking at their answering time, it seems that they did not have much time to complete the last questions, and therefore might have been in a rush and did not give correct answers. Finally, participants 20 to 24 seem to have tested completion, but preferred to finish the test without using it.

\begin{figure}
\centering
\includegraphics[width=\linewidth]{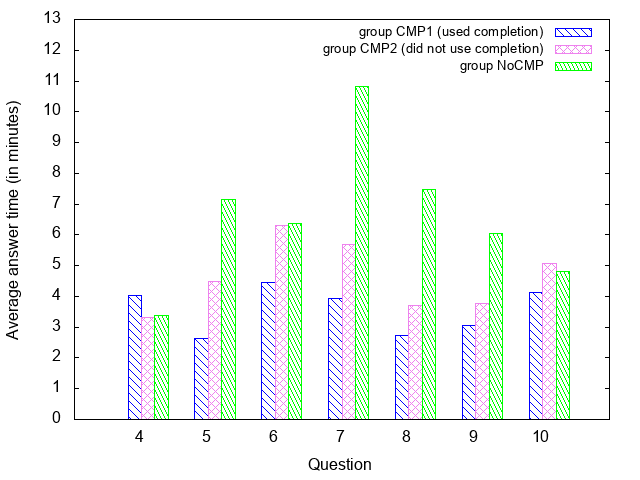}
\caption{Histogram of average answering time for questions 4 to 10, for groups CMP1, CMP2 and NoCMP}
\label{fig:g5}
\end{figure}

\begin{figure}
\centering
\includegraphics[width=\linewidth]{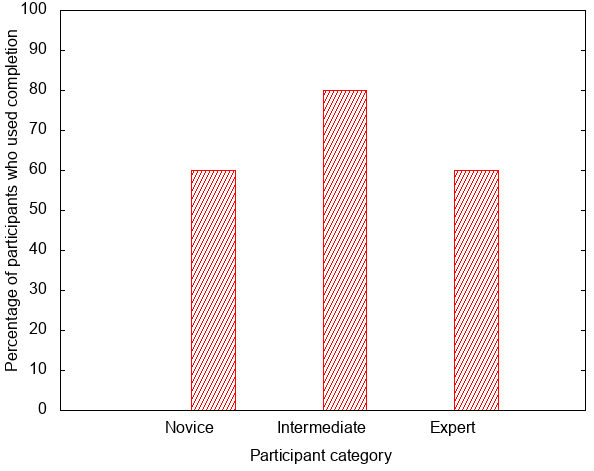}
\caption{Percentage of participants from group CMP who used completion, based on their estimated level in SQL}
\label{fig:g7}
\end{figure}

On figure \ref{fig:g5}, we divided group CMP into two groups for each question: participants who had submitted a query generated with completion (group CMP1), and others participants from group CMP in group CMP2. We then compared their average answering time for each questions, as well as for group NoCMP. It should be noticed that for each question groups CMP1 and CMP2 might be different as participant who used completion are different from one question to another. First, on question 4, group CMP1 is slower: as it is the first question on which completion could be used, we interpret this as the time necessary for participant to get familiar with the completion tool. Moreovoer, even though group CMP2 answered without completion, its behaviour is different from group NoCMP on question 4 to 10. Indeed, except for question 10 where it is the slowest group (but on previous figures, question 10 always has specific behaviours), the tendency of group CMP2 is closer to the one of group CMP1 than to the one of group NoCMP. 
This is explained by the fact that participants who did not use completion in group CMP were students good enough in SQL to be able to answer the question quickly: for them, taking the time to understand the completion tool would have been a waste of time as they were comfortable enough in SQLn an had enough information, to succeed the test without it. This correlates with figure \ref{fig:g7}: based on the participants study year and their self-estimated level in SQL, we divided group CMP into three categories of participants, novices, intermediates and experts. We then looked, for each category, what proportion of participants had used completion at least once. The results from figure \ref{fig:g7} show that the group that used it the most is the one for intermediate level. Expert used it less, because as we explained, they were comfortable enough in SQL to answer questions quickly. More surprisingly, the novice category does not seem to have used it more than experts: the explanation for this is found when looking at the queries submitted by participants from this group. They tried to find queries more complicated than they actually were, as they had a really scholar approach of the test.

To summarize, we believe those thorough experimentations have reach their goals.
We demonstrated that the group with the completion tool performed faster than the one without: on average, group CMP completed the test in 32 minutes, against 48 minutes for group NoCMP. This is not only because the tool allows to write faster, but mostly because it identifies conditions that take much more time to find manually, as it requires to try several values before finding the pertinent one.
We saw that the tool is well accepted, depending of the participant's context and level in SQL: moreover, we showed that the use of completion was not a single isolated try by participants, but that a first use encouraged them to use it again. \textit{Once you got it, you have it forever.}

\section{Related work}
\label{sec:state}

As far as we know, SQL query completion is a new problem that has not been studied yet. 
Nevertheless, related contributions exists in the context of data exploration.
The closest example is in \cite{khoussainova2010snipsuggest}, that proposes an autocompletion tool for SQL, that provides context aware assistance in SQL queries writing. Even though it can suggest completions in various SQL clauses, the completions offered rely mostly on the schema and on the database's log, and does not look at the data itself contrary to our approach that relies on the database's content.
Many approaches try to infer query based on \textit{example tuples}, which are tuples manually labelled by the user.
Among those we can cite \cite{Shen:2014:DQB:2588555.2593664} that suggest a set of queries returning such example tuples. Another example is \cite{Bonifati:2014:IJQ:2733004.2733025}, where the objective is to infer the join query that will return the result expected by the user. 
Those approach are similar to ours in the sense that their purpose if to help query formulation. However, the labelling done by the user is a additional task she has to do in addition to the usual ones. In comparison, we only ask an input query from the user.
The clustering phase of our solution is in charge of the labelling, which means many more tuples can be labelled as it is automatic. Moreover, it can lead the user to consider data she had not thought about before.

If we take the various component of the solution presented in this paper, they can be linked to several research areas.
First, \textit{Reverse query engineering} \cite{Tran:2014:QRE:2673202.2673266, zhang2013reverse}  considers the following problem: given a tuple set $T$ in a database $d$,  find a query $Q$ such that $ans(Q,d) = T$. Many theoretical results exist with respect to the language permitted to express $Q$, conjunctive queries and variants. This is what we do with the decision tree used to formulate a query returning the tuples from a cluster. However, in our context,  some simplification is permitted since part of the query is known. 
Second, \textit{Redesciption mining} \cite{parida2005redescription} unifies considerations of conceptual clustering, constructive induction, and logical formula discovery. Nevertheless, they do not consider at all SQL queries as we do, and are interested in enumerating all possible redescriptions verifying some conditions, with enumeration techniques quite different from our proposition.
Decision trees in databases have also been studied in various forms: in \cite{cumin2017data} they are used to reformulate in query for data exploration. We can also mention works on integrating decision trees into databases as objects that can be stored and queried, such as in \cite{fromont2006integrating} or \cite{rehman2010enabling}. Also related to our solution are predictive cluster trees that combine those two method into one \cite{liu2000clustering}.

Many recent works concern interactive data exploration in database, with techniques aiming at helping user understand and discover their data using machine learning. Some examples of such works are exposed in \cite{papaemmanouil2016interactive}. Many of those approaches also rely on manually labelled tuples : we can cite the \textit{AIDE} framework offers \cite{DBLP:journals/tkde/DimitriadouPD16} that tries to learn what tuples are of relevance for the user and which are not.
The machine learning phase is essentially based on decision trees and SVM. Thereafter, this process was improved in \cite{DBLP:journals/debu/PapaemmanouilDD16}, by using even more machine learning. This second paper uses the same framework, but completes it by identifying underlying user habits based on their labeling. Those habits are turned into attributes used in a clustering. This way, similar users are identified, which is used for speeding the process by using previous data exploration by similar users to give even more relevant tuples. We can see here many similar features with our query completion proposal. However once again those approaches require more work from the user. 

More generally, there is a part of research trying to bridge the gap between machine learning and database. Surajit Chaudhuri in \cite{Chaudhuri98datamining} argues that bringing databases and machine learning algorithms closer might only be beneficial in terms of performance. More concrete applications of this has actually been done, such as in \cite{Zou2006} where an entire machine learning library has been adapted so that it is compatible with a storage of data in a DBMS instead of a data structure in main memory. Moreover, they also adapted the algorithms in order to make use of native SQL operators.  
There is also \cite{wang2003atlas} which is a SQL extension for data mining.

In conclusion, we can see that this query completion framework is part of a general branch of research towards data exploration, which is motivated by the new challenges that \textit{Big Data} and the evolution of data science are bringing.

\section{Conclusion}
\label{sec:ccl}

In this paper, we have adapted for the first time the powerful notion of \emph{completion} to SQL queries, which could be particularly useful for data analysts in a data exploration process or for SQL developers. This functionality is a natural extension of SQL and could be integrated in every SQL editors associated to database management systems (DBMS).

Without any intervention required for the user, any SQL query can be completed automatically and should give rise to new ideas, new paths to the user in her quest to the elicitation of her data of interest.
The completion is semantic, and relies on the data contained in the answer set of this initial query. The approach is based on classical machine learning algorithms, adapted to fit into the definition of the completion we have proposed. 

A SQL editor prototype has been developed on top of which experimentations have been conducted over a set of 70 participants. It demonstrate the pertinence of such a tool in current DBMS: not only do participants get adapted to it easily, but it also allows to considerably improve and facilitate the SQL query writing process in the considered context. Contrary to syntactic completion tool, our approach does not only improve the writing itself, but it helps the analysts to identify data and to limit the number of iteration she has to do to identify relevant conditions to reach desired data.

In the current context where more and more data is being stored and analysed, such a proposal is a real help for data exploration, to assist analysts confronted to unknown databases, using completions to navigate and understand the data. Moreover, the solution is iterative and allows the user to modify a completion and to continue until she reaches what she was looking for. The completion is also a way to integrate knowledge on data, usually provided by data mining systems and tools, without leaving the context of DBMS.

Many extensions of this work can be envisioned, typically to extend the completion to different clauses of SQL, for instance the \texttt{group by} clause. This work is also a contribution to bridge the gap between database techniques and machine learning techniques.

\balance

\bibliographystyle{abbrv}
\bibliography{biblio}  

\begin{thebibliography}{10}

\bibitem{Bonifati:2014:IJQ:2733004.2733025}
A.~Bonifati, R.~Ciucanu, and S.~Staworko.
\newblock Interactive join query inference with jim.
\newblock {\em Proc. VLDB Endow.}, 7(13):1541--1544, Aug. 2014.

\bibitem{cart84}
L.~Breiman, J.~Friedman, R.~Olshen, and C.~Stone.
\newblock {\em {Classification and Regression Trees}}.
\newblock Wadsworth and Brooks, Monterey, CA, 1984.

\bibitem{breiman1984classification}
L.~Breiman, J.~Friedman, C.~J. Stone, and R.~A. Olshen.
\newblock {\em Classification and regression trees}.
\newblock CRC press, 1984.

\bibitem{conf/cidr/CetintemelCDDDKPZ13}
U.~{\c{C}}etintemel, M.~Cherniack, J.~DeBrabant, Y.~Diao, K.~Dimitriadou,
  A.~Kalinin, O.~Papaemmanouil, and S.~B. Zdonik.
\newblock Query steering for interactive data exploration.
\newblock In {\em {CIDR} 2013, Sixth Biennial Conference on Innovative Data
  Systems Research, Asilomar, CA, USA, January 6-9, 2013, Online Proceedings},
  2013.

\bibitem{Chaudhuri98datamining}
S.~Chaudhuri.
\newblock Data mining and database systems: Where is the intersection?
\newblock {\em Data Engineering Bulletin}, 21, 1998.

\bibitem{cumin2017data}
J.~Cumin, J.-M. Petit, V.-M. Scuturici, and S.~Surdu.
\newblock Data exploration with sql using machine learning techniques.
\newblock In {\em International Conference on Extending Database
  Technology-EDBT}, 2017.

\bibitem{DBLP:journals/tkde/DimitriadouPD16}
K.~Dimitriadou, O.~Papaemmanouil, and Y.~Diao.
\newblock {AIDE:} an active learning-based approach for interactive data
  exploration.
\newblock {\em {IEEE} Trans. Knowl. Data Eng.}, 28(11):2842--2856, 2016.

\bibitem{fromont2006integrating}
{\'E}.~Fromont, H.~Blockeel, and J.~Struyf.
\newblock Integrating decision tree learning into inductive databases.
\newblock In {\em International Workshop on Knowledge Discovery in Inductive
  Databases}, pages 81--96. Springer, 2006.

\bibitem{Han:2005:DMC:1076797}
J.~Han.
\newblock {\em Data Mining: Concepts and Techniques}.
\newblock Morgan Kaufmann Publishers Inc., San Francisco, CA, USA, 2005.

\bibitem{khoussainova2010snipsuggest}
N.~Khoussainova, Y.~Kwon, M.~Balazinska, and D.~Suciu.
\newblock Snipsuggest: Context-aware autocompletion for sql.
\newblock {\em Proceedings of the VLDB Endowment}, 4(1):22--33, 2010.

\bibitem{Levene:1999:GTR:553537}
M.~Levene and G.~Loizou.
\newblock {\em A Guided Tour of Relational Databases and Beyond}.
\newblock Springer-Verlag, London, UK, UK, 1999.

\bibitem{liu2000clustering}
B.~Liu, Y.~Xia, and P.~S. Yu.
\newblock Clustering through decision tree construction.
\newblock In {\em Proceedings of the ninth international conference on
  Information and knowledge management}, pages 20--29. ACM, 2000.

\bibitem{Lloyd:2006:LSQ:2263356.2269955}
S.~Lloyd.
\newblock Least squares quantization in pcm.
\newblock {\em IEEE Trans. Inf. Theor.}, 28(2):129--137, Sept. 2006.

\bibitem{journals/pvldb/NandiJ11}
A.~Nandi and H.~V. Jagadish.
\newblock Guided interaction: Rethinking the query-result paradigm.
\newblock {\em PVLDB}, 4(12):1466--1469, 2011.

\bibitem{papaemmanouil2016interactive}
O.~Papaemmanouil, Y.~Diao, K.~Dimitriadou, and L.~Peng.
\newblock Interactive data exploration via machine learning models.
\newblock {\em IEEE Data Eng. Bull.}, 39(4):38--49, 2016.

\bibitem{DBLP:journals/debu/PapaemmanouilDD16}
O.~Papaemmanouil, Y.~Diao, K.~Dimitriadou, and L.~Peng.
\newblock Interactive data exploration via machine learning models.
\newblock {\em {IEEE} Data Eng. Bull.}, 39(4):38--49, 2016.

\bibitem{parida2005redescription}
L.~Parida and N.~Ramakrishnan.
\newblock Redescription mining: Structure theory and algorithms.
\newblock In {\em AAAI}, volume~5, pages 837--844, 2005.

\bibitem{scikit-learn}
F.~Pedregosa, G.~Varoquaux, A.~Gramfort, V.~Michel, B.~Thirion, O.~Grisel,
  M.~Blondel, P.~Prettenhofer, R.~Weiss, V.~Dubourg, J.~Vanderplas, A.~Passos,
  D.~Cournapeau, M.~Brucher, M.~Perrot, and E.~Duchesnay.
\newblock Scikit-learn: Machine learning in {P}ython.
\newblock {\em Journal of Machine Learning Research}, 12:2825--2830, 2011.

\bibitem{rehman2010enabling}
N.~U. Rehman and M.~H. Scholl.
\newblock Enabling decision tree classification in database systems through
  pre-computation.
\newblock In {\em British National Conference on Databases}, pages 118--121.
  Springer, 2010.

\bibitem{rushby2016wiley}
N.~Rushby and D.~Surry.
\newblock {\em The Wiley Handbook of Learning Technology}.
\newblock Wiley Handbooks in Education. Wiley, 2016.

\bibitem{Shen:2014:DQB:2588555.2593664}
Y.~Shen, K.~Chakrabarti, S.~Chaudhuri, B.~Ding, and L.~Novik.
\newblock Discovering queries based on example tuples.
\newblock In {\em Proceedings of the 2014 ACM SIGMOD International Conference
  on Management of Data}, SIGMOD '14, pages 493--504, New York, NY, USA, 2014.
  ACM.

\bibitem{Tran:2014:QRE:2673202.2673266}
Q.~T. Tran, C.-Y. Chan, and S.~Parthasarathy.
\newblock Query reverse engineering.
\newblock {\em The VLDB Journal}, 23(5):721--746, Oct. 2014.

\bibitem{wang2003atlas}
H.~Wang, C.~Zaniolo, and C.~R. Luo.
\newblock Atlas: A small but complete sql extension for data mining and data
  streams.
\newblock In {\em Proceedings of the 29th international conference on Very
  large data bases-Volume 29}, pages 1113--1116. VLDB Endowment, 2003.

\bibitem{wu2016decision}
C.-C. Wu, Y.-L. Chen, Y.-H. Liu, and X.-Y. Yang.
\newblock Decision tree induction with a constrained number of leaf nodes.
\newblock {\em Applied Intelligence}, 45(3):673--685, 2016.

\bibitem{zhang2013reverse}
M.~Zhang, H.~Elmeleegy, C.~M. Procopiuc, and D.~Srivastava.
\newblock Reverse engineering complex join queries.
\newblock In {\em Proceedings of the 2013 ACM SIGMOD International Conference
  on Management of Data}, pages 809--820. ACM, 2013.

\bibitem{Zou2006}
B.~Zou, X.~Ma, B.~Kemme, G.~Newton, and D.~Precup.
\newblock {\em Data Mining Using Relational Database Management Systems}, pages
  657--667.
\newblock Springer Berlin Heidelberg, Berlin, Heidelberg, 2006.

\end{thebibliography}

\newpage


\begin{appendix}
\label{app:test}
The presentation of the database used for our experimentations, as well as the questions asked to the participants, are presented in this appendix. 

\section{Experimentation Scenario}
You're a new member of a post office, in charge of packages. When you're not at the front desk taking care of customers, you have access to data recorded about the packages sent from your post office. For simplification, we will focus on the packages leaving the post office to other destinations. Here is how the database was created :

\begin{lstlisting}[basicstyle=\small]
CREATE TABLE Cities(
   city_ID DECIMAL,
   distance DECIMAL,
   PRIMARY KEY (city_ID)
)
CREATE TABLE Packages(
    package_ID DECIMAL,
    destination DECIMAL,
    length DECIMAL,
    width DECIMAL,
    height DECIMAL,
    weight DECIMAL,
    price DECIMAL,
    PRIMARY KEY (id_colis)
    FOREIGN KEY (destination) 
    	references Villes(id_ville)
)
\end{lstlisting}

Table Packages has one entry per package that left your post office. From the destination of a package, you can see how far it was sent, by joining tables \textit{Packages} (11000 tuples) and \textit{Cities} (30 tuples) on attributes \textit{destination} and \textit{city\_ID}.


Questions are ordered from easiest to hardest :you should therefore answer them in the given order. First three questions are simple, while the others are voluntary more complex, and finding the required SQL query in questions 4 to 10 required more exploration.

\section{Questions}

\textbf{Question 1}
This first question is here so that you can get familiar with the data and the tools at your disposal. Please test the two tools (SQL software and online form for answers) with the following query, that is a join between the two tables (Expected result size: 10 999 tuples):

\begin{lstlisting}[basicstyle=\small]
Select *
From Packages, Cities
Where Packages.destination = Cities.city_ID
\end{lstlisting}

\textbf{Question 2}
Maximum size limit authorized for a package is 9000 grams. However, some exceed this limit without being detected. Give the query to obtain the ID of packages whose weight exceed this limit. (Expected result size:  73 tuples)

\textbf{Question 3}
What query can you write to obtain the average length of packages sent less than 100 kilometers from your post office ? (Expected result size: 1 tuple)

\begin{figure}[H]
\centering
\includegraphics[width=.9\linewidth]{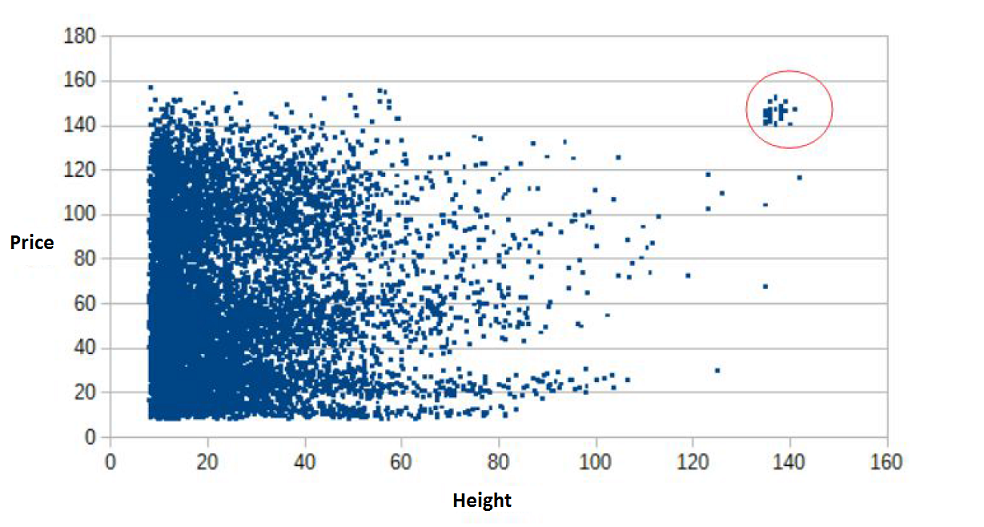}
\caption{Visualization for question 4}
\label{fig:q4}
\end{figure}

\textbf{Question 4}
A little bit interested by data analysis, a colleague of yours had, with a spreadsheet, visualized some curves from the database. 
By plotting packages prices against their height, he/she had noticed a group of packages very distinct and well separated from the others, which is presented on figure \ref{fig:q4}, and circled in red.
Can you find the query that returns all packages belonging to this group ?(Expected result size: 33 tuples)

\textbf{Question 5}
According to some colleagues who've been working here for years, heaviest packages are the one going to very distant destinations. The intuition behind this is that has sending a package far away is expansive, customers many things in one package to compensate.
Can you identify packages that do not comply with this, i.e that are not heavy but are sent far away ? (Expected result size: 13 tuples)

\textbf{Question 6}
Once at the regional sorting center, packages go through a machine that automatically sort them according to their destination. However, this machine is sometimes defective. Indeed, when a package is less than 480g, the machine does not always detect it, and a operator has to take it and process it manually. This phenomenon is marginal, but more likely to happen if in addition to its light weight, the packages in small regarding its length and width. On all packages registered in your database, 12 have caused a problem. 
Which query can identify those 12 packages ?

\textbf{Question 7}
Some packages are sent to a city that is very close to your post office, less than 10km away. Moreover, some are very light (less than 550g), and you wonder why people pay the post office to transport them while they would quite easily do it themselves.
One of your colleagues has an hypothesis : maybe those packages are cumbersome and therefore hard to transport. 
Can you identify packages validating this hypothesis ?
(Expected result size: 8 tuples)

\textbf{Question 8}
A customer arrives at the post office, because he needs the ID of a package he had send, but isn't able to find. In order to help him, he gives you a few informations: the package was light, less than 450g and its dimensions (mainly length and width) were surprisingly big in regard to its weight.
Can you give the query returning such a package ? (Expected result size: 1 tuple)

\textbf{Question 9}
When working at the front desk, one of your colleagues made a mistakes on four on the packages he registered. Luckily, he remembers their length was above 140cm, and he therefore applied a special tarification, as those kind of packages are more complicated to deliver due to their size. But he applied the wrong tarification, and those packages have therefore an abnormally elevated price.
Can you identify those packages ?
(Expected result size: 4)

\textbf{Question 10}
At question 2, you showed that 73 packages are above the weight limit. But your colleagues in charge of putting packages in the trucks say that a third of packages are really heavy, and require two employees to be lifted, in order to avoid back pains.
Can you modify the query for question 2 in order to identify those packages ?
(Expected result size : 3073 tuples)
\end{appendix}

\end{document}